\documentclass[12p]{iopart}

\usepackage{color}
\usepackage{graphicx}
\usepackage{iopams}
\usepackage{braket}
\usepackage[utf8]{inputenc}
\usepackage{url}
\usepackage{bbold}
\usepackage{multirow}

\newcommand{\Fig}[1]{Fig.~\ref{#1}}

\begin{document}

\title[Direct generation of genuine single-longitudinal-mode narrowband photon pairs]
{Direct generation of genuine single-longitudinal-mode narrowband photon pairs}

\author{Kai-Hong Luo, Harald Herrmann, Stephan Krapick, Benjamin Brecht, Raimund Ricken, Viktor Quiring, Hubertus Suche, Wolfgang Sohler, and Christine Silberhorn}

\address{Integrated Quantum Optics, Applied Physics, University of Paderborn,
Warburger Str. 100, 33098, Paderborn, Germany}

\ead{khluo@mail.uni-paderborn.de}

\date{\today}

\begin{abstract}
The practical prospect of quantum communication and information processing relies on sophisticated single photon pairs which feature controllable waveform, narrow spectrum, excellent purity, fiber compatibility and miniaturized design. For practical realizations, stable,
miniaturized, low-cost devices are required. Sources with one or some of above performances have been demonstrated already, but it is quite challenging to have a source with all of the described characteristics simultaneously. Here we report on an integrated single-longitudinal-mode non-degenerate narrowband photon pair source, which exhibits all requirements needed for quantum applications. The device is composed of a periodically poled Ti-indiffused lithium niobate waveguide with
high reflective dielectric mirror coatings deposited
on the waveguide end-faces. Photon pairs with wavelengths around 890 nm and 1320 nm are generated via type II phase-matched parametric down-conversion. Clustering in this dispersive cavity restricts the whole conversion spectrum to one single-longitudinal-mode in a single cluster yielding a narrow bandwidth of only 60 MHz. The high conversion efficiency in the waveguide, together with the spectral clustering in the doubly resonant waveguide, leads to a high brightness of $3\times10^4~$pairs/(s$\cdot$mW$\cdot$MHz). This source exhibits prominent single-longitudinal-mode purity and remarkable temporal shaping capability. Especially, due to temporal broadening, we can observe that the coherence time of the two-photon component of PDC state is actually longer than the one of the single photon states. The miniaturized monolithic design makes this source have various fiber communication applications.
\end{abstract}


\maketitle

\section{Introduction}

Quantum communication and information processing (QCIP)\cite{GisinRMP2002,GisinNP2007} currently evolve from fundamental research towards real life applications. This process can be strongly fostered by the implementation of integrated optical devices offering miniaturized and potentially low-cost components for applications in quantum key distribution \cite{GisinRMP2002}, long distance quantum communication \cite{DuanN2001}, quantum repeaters \cite{SangouardRMP2011}, and quantum networks \cite{KimbleN2008}. Moreover, compact and rugged integrated optical quantum circuits \cite{PolitiAS2008,SansoniLPRL2010,TanzilliLPR2012,OBrienNJP2013,SpringS2013} with high functionality can efficiently replace bulky implementations to pave the way to practical applications.

In particular, quantum repeater architectures have been proposed \cite{SangouardRMP2011,BriegelPRL1998,SimonPRL2007} to overcome current limitations of long distance quantum communication due to transmission losses.
These typically require spectrally narrowband two-color photon pairs for instance to address the absorption line of the storage medium in a quantum memory (QM) \cite{JulsgaardNat2004,LvovskyNP2009,WalmsleyNP2010,LamNP2011,ZhangNP2011,TittelJMO2013} with one photon and transmit the second over a fiber network. Such QMs usually have their absorption in the visible or near infrared, i.e.\ far away from the telecommunication range. The corresponding bandwidths are typically in the range of MHz to GHz, depending on different storage regimes (like spin-wave storage in cold atomic ensembles, atomic-frequency-comb echoes in solid state and room-temperature storage ). Among the most promising materials for high-bandwidth QM's are solid-state atomic ensembles, specifically rare-earth ion doped crystals or glasses \cite{SellarsLPR2010,TittelN2011,GisinN2011,RiedmattenPRL2014} and QM at room-temperature \cite{Michelberger2014,EnglandPRL2015}.

A wide-spread method to generate photon pairs is parametric down conversion
(PDC) \cite{BurnhamPRL1970}. In such a PDC process, a medium with $\chi^{(2)}$ nonlinearity splits a single pump photon into two photons of lower energy, named  signal and  idler, obeying energy conservation and phase matching. However, the loose phase-matching condition for both, bulk and waveguide sources, usually leads to a continuous broad bandwidth typically exceeding several 100 GHz and a mixed state in frequency. To overcome this bottle-neck, narrowband photon-pair sources are desired with an adapted bandwidth and a high spectral brightness.

One promising approach to generate such narrowband photon pairs is to
use resonance enhancement of PDC within a cavity, also called
optical parametric oscillator (OPO) far below the threshold \cite{OuPRL1999,LauratPRA2006,
KuklewiczPRL2006, PanPRL2008, BensonPRL2009, HockelPRA2011, GuoOL2008,PolzikOL2009,
WolfgrammPRL2011, PomaricoNJP2009, PomaricoNJP2012,
URenLP2010, ChuuAPL2012, FortschNC2013, FeketePRL2013,FortschPRL2014}. PDC is enhanced at the resonances of the cavity but inhibited at non-resonant frequencies. An overview over the most important resonant PDC sources published so far it given in Tab.~\ref{tb:RefResult01}. Despite an enormous progress -- non-degenerate photon pairs
with linewidths of about 2 MHz oscillating on 4 longitudinal modes -- which was obtained in the
bulk cavity \cite{FeketePRL2013}, most of the bulk sources suffer from low pair production efficiency, degenerate frequency, bad mode selectivity and complicated cavity locking techniques.

To overcome these limitations strong benefits from a monolithic implementation of this scheme can be expected. Recently, a high Q cavity using a whispering gallery mode resonator has been demonstrated to achieve single-mode photon pairs \cite{FortschPRL2014}.
An alternative approach using a periodically poled lithium niobate (PPLN) waveguide with dielectric mirrors deposited on its end-faces has been demonstrated to generate nearly degenerate photon-pairs via a type I phase-matched PDC process in the telecom region\cite{PomaricoNJP2009}. However,
it exhibited a limited mode selectivity, because the spectrum consisted of a series of longitudinal modes due to the small difference of the free spectral ranges (FSRs) of signal and idler close to degeneracy. A detailed theoretical study \cite{PomaricoNJP2012} indicates that exploiting type II phase-matching should strongly limit the number of longitudinal modes because the large birefringence provides a larger difference of the FSRs, thus, fosters the spectral narrowing.

\begin{table*}[hbp]\scriptsize
\centering
\label{tb:RefResult01}
\begin{tabular}{|c|c|c|c|c|c|}
\hline
\multirow{2}{*}{PDC} & Bao \textit{et al} & Scholz \textit{et al} & Chuu \textit{et al} & Fekete \textit{et al} & {Ours}\\
 & \cite{PanPRL2008} & \cite{BensonPRL2009} & \cite{ChuuAPL2012} & \cite{FeketePRL2013} &\\
\hline
{wavelength (nm)} & 780 & 893 & 1064 & 606+1436 & 890+1320\\
\hline
{bandwidth (MHz)} & 9.6 & 2.7 & 8.3 & 1.7 & 60\\
\hline
{brightness ($\frac{pairs}{s \cdot mW  \cdot MHz)}$} & 6 & 330 & $1.34 \times 10^4$ & 11 & $3\times10^4$\\
\hline
{mode selectivity} & multi-mode & single-mode & -- &several modes & single-mode\\
\hline
{fiber compatibility} &  -- &  -- & \checkmark & \checkmark & \checkmark\\
\hline
{atom interaction} &  Rb (MHz) &  Cs (MHz) & -- & Pr (MHz)  & Nd (~100 MHz) \\
\hline
{cavity locking} &  single cavity &  double cavity & no & double cavity & no\\
\hline

\end{tabular}
\caption{Recent results of cavity-enhanced PDC are shown in the table. All sources except ours described in the last column used bulk crystals to generate narrowband photon pairs, which either feature single-mode operation, or have high brightness, or combine fiber network with quantum memory. For comparison, the last column shows the results of the integrated source reported in this paper.}
\end{table*}

The drawback of the resonant sources reported so far is that they either achieve narrow bandwidth, or operate single mode, or reach high brightness, or combine two wavelengths for atomic transition and fiber communications. However, for practical quantum application, sources which provide simultaneously all the above mentioned properties are required. In this paper, we present the experimental realization of such a miniaturized two-color integrated bright narrowband photon pair source based on a doubly resonant waveguide exploiting type II phase-matching exhibiting one longitudinal mode.

In the Sec.\ref{theory} we will discuss in detail the principle of operation and the underlying physical theory. In the non-degenerate case, birefringence and material dispersion result in different FSRs for signal and idler. As maximum efficiency is only obtained if both signal and idler are resonant simultaneously, PDC is generated only in certain regions of the spectrum, so called 'clusters' \cite{EckardtJOSAB1991}. In this way the spectral density is redistributed in comparison to the non-resonant case and, thus, ideally a completely filter-free source with actively reduced bandwidth can be realized without sacrificing any photon flux level. In Sec.\ref{Fabrication} and Sec.\ref{Setup} details on the waveguide fabrication and the experimental set-up are given, respectively. The practical implementation is challenging because it requires loss minimization, short poling periods for phase-matching and specifically tailored mirror coatings for both signal and idler photons. In Sec.\ref{Results} the experimental results are discussed and compared with the theoretically predicted ones. Our engineered clustering in the dispersive cavity restricts the conversion to a single longitudinal mode operation with a spectral linewidth around 60 MHz and high spectral brightness $3\times10^4~$pairs/(s$\cdot$mW$\cdot$MHz). Such a non-degenerate PDC photon pair source can be applied in a quantum repeater addressing a Nd-based QM with one photon and combining the other with a fiber network.

\begin{figure}
\begin{center}
\includegraphics*[width=0.9\textwidth]{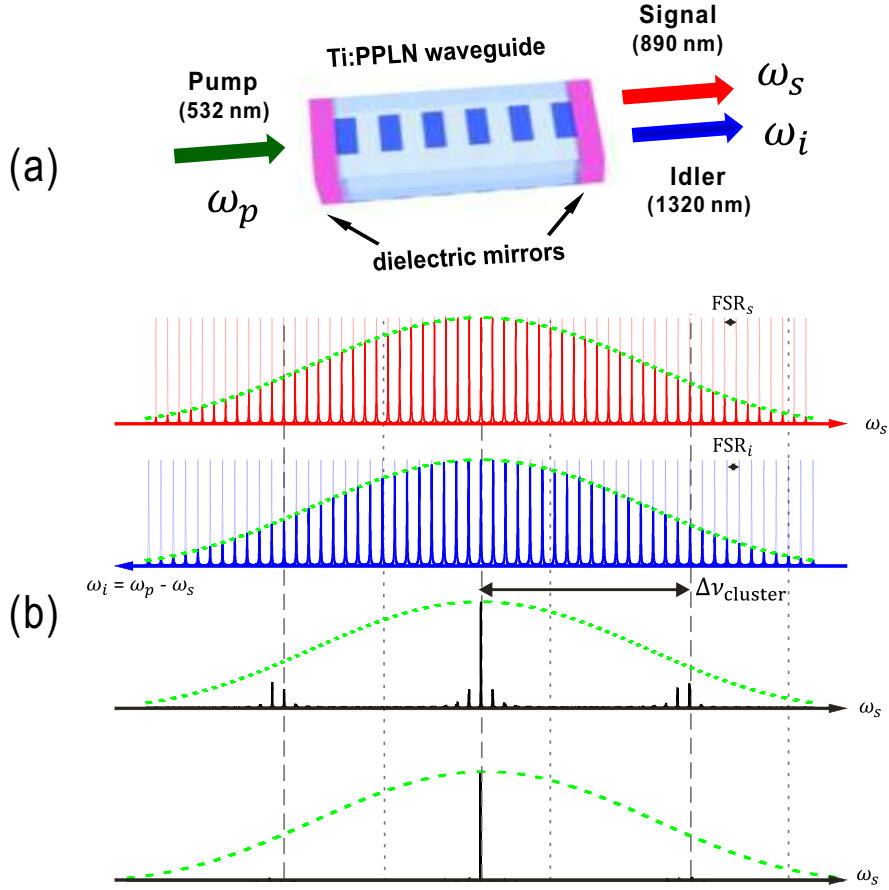}
\end{center}
\caption{\label{Resonantcluster} An integrated compact narrowband photon pair source in a resonant waveguide and its working principle.
(a) Integrated doubly resonant waveguide. The source consists of a Ti-indiffused periodically poled lithium niobate (Ti:PPLN) waveguide and dielectric mirrors deposited on the waveguide end-faces. The first-order type II phase-matching was chosen to generate TE-polarized signal photons around 890~nm (red) and TM-polarized photons around 1320~nm (blue) when pumped at $\lambda_p= $532~nm in TE-polarization.
(b) Clustering effect.
The cavity resonances form frequency combs at the signal and idler wavelength range (red and blue) with different free spectral ranges ($\rm{FSR}_{s,i}$) due to the dispersion in the waveguide. Photon pair generation by PDC obeying energy conservation $\omega_i=\omega_p-\omega_s$ and phase-matching (with a spectral envelope indicated by the green dashed curve) occurs only if signal and idler are resonant simultaneously. Thus, the product of the two frequency combs and the phase-matching envelope (black solid lines) determines the spectrum of the resonant PDC source with cluster separation $\Delta \nu_{\textrm{cluster}}$. By increasing the finesse of cavity (i.e. narrowing the bandwidth of the individual comb lines), the side clusters and modes can be suppressed (bottom black one), to achieve a single longitudinal mode operation.}
\end{figure}

\section{Principle and theory}\label{theory}
\subsection{Nonresonant PDC process}
PDC is an important process in quantum optics, used especially as
a source of entangled photon pairs and of single photons. In a nonlinear $\chi ^{(2)}$ medium
 pump photons (at frequency $\omega_p$) are split into pairs of photons
(signal $\omega_s$ and idler $\omega_i$, respectively)  obeying
energy and momentum conservation. As for most cases the latter cannot be obtained in homogeneous crystals, quasi-phase-matching is
exploited in a periodically inverted medium. Thus, the phase-matching condition becomes:
\begin{equation}
\Delta k  = {k\left( {\omega _p} \right) - k\left( {{\omega _s}} \right) - k\left( {{\omega _i}} \right)}- \frac{2\pi}{\Lambda}\approx 0
\end{equation}
with $k_j~(j=p,s,i)$ are the propagation constants of the pump, signal and idler fields and $\Lambda$ is the poling period. We express the state of the photon pairs generated by PDC up to second order expansion,
\begin{eqnarray}
&&{\left| \Psi  \right\rangle _{PDC}} \\ \nonumber
& \propto &
 \int d{\omega _s}d{\omega _i}{f \left( {{\omega _s}{\rm{,}}{\omega _i}} \right)} \hat a_s^\dag \left( {{\omega _s}} \right)\hat a_i^\dag \left( {{\omega _i}} \right)\left| 0 \right\rangle \\ \nonumber
& + &\int d{\omega _s}d{\omega _s'}d{\omega _i}d{\omega _i'}{f \left( {{\omega _s}{\rm{,}}{\omega _i}} \right)}{f \left( {{\omega _s'}{\rm{,}}{\omega _i'}} \right)} \hat a_s^\dag \left( {{\omega _s}} \right)a_s^\dag \left( {{\omega _s'}} \right)\hat a_i^\dag \left( {{\omega _i}} \right)\hat a_i^\dag \left( {{\omega _i'}} \right)\left| 0 \right\rangle,
\label{eq:PDCstate0}
\end{eqnarray}
where ${{\hat a^\dag}_{s,i}}\left( {{\omega _{s,i}}} \right)$ describes the photon creation operator at
frequency ${{\omega _{s,i}}}$ and  ${f\left( {{\omega _s}{\rm{,}}{\omega _i}} \right)}$ is the joint spectral function (JSF) of the nonlinear waveguide:
\begin{eqnarray}
{f\left( {{\omega _s}{\rm{,}}{\omega _i}} \right)}
= \alpha \left( {{\omega _s}+{\omega _i}} \right)\times {\phi\left( {{\omega _s}{\rm{,}}{\omega _i}} \right)}.
\label{eq:JSA0}
\end{eqnarray}
The $\alpha \left( {{\omega _s}+{\omega _i}} \right)$ describes the spectrum and amplitude of the pump
field, ${\phi\left( {{\omega _s}{\rm{,}}{\omega _i}} \right)}$ is phase-matching function ${\rm {sinc}} \left[ {\Delta k \left( {{\omega _s}{\rm{,}}{\omega _i}} \right)\frac{L}{2}} \right]{\rm {exp}} \left[ i{\Delta k \left( {{\omega _s}{\rm{,}}{\omega _i}} \right)\frac{L}{2}} \right]$.

The full-width at half-maximum (FWHM) of the PDC process in the waveguide is determined by the phase-matching condition. Assuming
a monochromatic pump, i.e. $\alpha \left( {{\omega _s}+{\omega _i}} \right)\propto\delta (\omega_s + \omega_i)$,  $\Delta k$ can be linearized
\begin{eqnarray}
\Delta k  ({\omega _{s,i}}) \approx \Delta {k_0} + \frac{1}{c}\left[ {{n_{gs}} - {n_{gi}}}\right]\Delta {\omega _{s,i}},
\label{eq:TaylorDk1}
\end{eqnarray}
where ${n_{gs}}$ and ${n_{gi}}$ are the group index of signal and idler, respectively. The spectral characteristics of the PDC process is given by a $\rm{sinc}^2\left[ {\Delta k ({\omega _s,\omega _i}) \frac{L}{2}} \right]$ function. Thus, the FWHM of the signal photons generated in a waveguide of length $L$ is given by:
\begin{eqnarray}
\Delta {\nu _s}^{\rm{FWHM}} & \approx & 5.56\frac{c}{{2\pi L}}\frac{1}{{\left| {{n_{gs}} - {n_{gi}}} \right|}},
\label{eq:SignalFWHM1}
\end{eqnarray}
i.e. the spectral bandwidth is inversely proportional to the difference of the group velocities of signal and idler. Thus, close to degeneracy the bandwidth is very large, whereas group velocity dispersion in most cases leads to a much narrower bandwidth far away from degeneracy.

\subsection{Resonant PDC}
Our source exploits photon pair generation by PDC within a cavity.
The schematic of the integrated source is shown in \Fig{Resonantcluster} (a). The resonant waveguide is a Fabry-P\'{e}rot cavity with asymmetric reflectivities ($R_{1s,i}$ and $R_{2s,i}$),
internal loss ($\alpha_{s,i}$) and effective indexes $n(\omega_{s,i})$ for the fundamental modes at signal and idler wavelength, respectively.
The photon pair state generated within such a resonator can be expressed by:
\begin{eqnarray}
\label{eq:RPDCstate1}
&&{\left| \Psi  \right\rangle _{RPDC}} \\ \nonumber
& \propto &
 \int d{\omega _s}d{\omega _i}{f_R \left( {{\omega _s}{\rm{,}}{\omega _i}} \right)} \hat a_s^\dag \left( {{\omega _s}} \right)\hat a_i^\dag \left( {{\omega _i}} \right)\left| 0 \right\rangle \\ \nonumber
& + &\int d{\omega _s}d{\omega _s'}d{\omega _i}d{\omega _i'}{f_R \left( {{\omega _s}{\rm{,}}{\omega _i}} \right)}{f_R \left( {{\omega _s'}{\rm{,}}{\omega _i'}} \right)} \hat a_s^\dag \left( {{\omega _s}} \right)a_s^\dag \left( {{\omega _s'}} \right)\hat a_i^\dag \left( {{\omega _i}} \right)\hat a_i^\dag \left( {{\omega _i'}} \right)\left| 0 \right\rangle,
\end{eqnarray}
where ${f_R \left( {{\omega _s}{\rm{,}}{\omega _i}} \right)}$ is the cavity-modified JSF:
${f_R \left( {{\omega _s}{\rm{,}}{\omega _i}} \right)}=f \left( {\omega _s}{\rm{,}}{\omega _i} \right) A_s\left( {\omega _s} \right)A_i\left( {\omega _i} \right)$, with ${f\left( {{\omega _s}{\rm{,}}{\omega _i}} \right)}$ being the conventional JSF for the nonresonant PDC. $A_{s,i}({\omega _{s,i}} )$ are proportional to the field distributions inside the cavity. They can easily be derived treating the device as a classical Fabry-P\'{e}rot resonator with internal loss:
\begin{eqnarray}
{A_{s,i}}\left( {{\omega _{s,i}}} \right) &= &{\sqrt {\left( {1 -R_{1s,i}} \right)\left( {1 -R_{2s,i}} \right)} e^{ - {\alpha _{s,i}}L/2}} \over
{1 - \sqrt {R_{1s,i}R_{2s,i}} e^{ - {\alpha _{s,i}}L}e^{i2{\omega _{s,i}}n\left( {\omega _{s,i}} \right)L/c}}.
\label{eq:CavAsAi2}
\end{eqnarray}
Therefore, we obtain an expression for the resulting
joint spectral intensity,
\begin{eqnarray}
S_{R} \left( {{\omega _s}{\rm{,}}{\omega _i}} \right) = |f_R ( {{\omega _s}{\rm{,}}{\omega _i}})|^2=|f ( {{\omega _s}{\rm{,}}{\omega _i}})|^2 \mathcal{A}_s \left( {\omega _s}\right)
\mathcal{A}_i\left({\omega _i} \right),
\label{eq:CavJSI1}
\end{eqnarray}
where $\mathcal{A}_{s,i} \left( {\omega _{s,i}}\right) = |A_{s,i} \left( {\omega _{s,i}}\right)|^{2}$
are Airy functions of signal and idler photons, respectively. The bandwidth of each cavity mode is related to the finesse, more details in \ref{sec:appendixfinesse}. Therefore the spectral characteristics is a product of phase-matching sinc function and two combs of Lorentzian lines, spaced by the resonator's FSRs.

The explicit form of signal spectrum can be determined completely as
\begin{eqnarray}
{S_{R}(\omega_s)} & \propto & \left| \int d{\omega_i}{f_R \left( {{\omega _s}{\rm{,}}{ \omega _i}} \right)} \right|^2\\ \nonumber
 & \propto & \rm{sinc}^2 \left[ \Delta k ({\omega _s}) \frac{L}{2} \right] \mathcal{A}_s \left( {\omega _s}\right)
\mathcal{A}_i\left({\omega _p -\omega _s} \right).
\label{eq:OutSpectra1}
\end{eqnarray}

\subsection{Cluster effect}

In the dispersive waveguide, cavity resonances occur at distinct frequencies separated by the free spectral range (FSR) of the resonator. These resonances form frequency combs spaced with respective FSRs in the signal and idler wavelength range (as shown in the Supplementary Movie).
\begin{eqnarray}
{\rm{FSR}}_{s,i} \approx \frac{c}{{2{n_{g{s,i}}} L}},
\label{eq:FSRdefinition1}
\end{eqnarray}
The FSR is the inverse of the round-trip time of a photon in the cavity.

The FWHM of each resonance , $\Delta \nu$, is related with the finesse $\mathcal{F}$ of the cavity:
\begin{eqnarray}
\mathcal{F}_{s,i}  =  \frac {{\rm{FSR}}_{s,i}}{\Delta \nu _{s,i}}
 =  \frac{{\pi \sqrt {\sqrt {{R_{1_{s,i}}}{R_{2_{s,i}}}} {e^{ - \alpha_{s,i} L}}} }}
{{\left( {1 - \sqrt {{R_{1_{s,i}}}{R_{2_{s,i}}}} {e^{ - \alpha_{s,i} L}}} \right)}}.
\label{eq:finessedefinition1}
\end{eqnarray}
For the type I degenerate case, the frequency combs of signal and idler photons are identical. Thus, PDC generation still happens at all comb frequencies within the phase-matching bandwidth. However, for type II and/or degenerate case,
there are different FSRs in the resonant waveguide due to the different group dispersions at the signal and idler wavelengths, as shown in the upper two curves in \Fig{Resonantcluster}(b). Thus, corresponding resonances of signal and idler only overlap at a certain frequency, but adjacent modes do not coincide,
as shown in the bottom curve in \Fig{Resonantcluster}(b). For instance, if a pair is simultaneously resonant for e.g.\ $\omega_{s0}$ and $\omega_{i0}$,
the adjacent signal resonance lies at $\omega_{s0}+2\pi{\rm{FSR}}_s$, and the corresponding idler frequency at $\omega_{i0}-2\pi{\rm{FSR}}_s$ is not in the resonance peak due to the different FSRs.
The couple of simultaneous resonances for
signal and idler is called a cluster \cite{EckardtJOSAB1991,PomaricoNJP2012}. However, after a certain
number of free spectral ranges, another cluster occurs again, if the phase-matching bandwidth is broad enough.
This happens when $N_0$ times the signal FSR equals $N_0-1$ times the idler FSR.
Thus, the cluster separation $\Delta \nu_{\textrm{cluster}}$, i.e.\ the frequency spacing between clusters, is given by:
\begin{eqnarray}
\Delta \nu_{\textrm{cluster}} = N_0 \cdot {\rm{FSR}}_s = \frac{1}{|{{n_{gi}} - {n_{gs}}}|}\frac{c}{{2L}},
\label{eq:Cluasterseparation2}
\end{eqnarray}
which is associated with the difference of the group indices. By comparing Eqs.~(\ref{eq:SignalFWHM1}) and (\ref{eq:Cluasterseparation2}), we can see that the cluster separation $\Delta {\nu_{\textrm{cluster}}}$ is slightly larger than half of the bandwidth of the PDC phase-matching envelope $\Delta {\nu _s}^{\rm{FWHM}}$. Therefore, the maximum number of clusters within the phase-matching range is 3.  Theoretically, if there is a dominant cluster in the center of the normalized phase-matching curve, two symmetric clusters will occur at the side wings of the envelope with about 41\% weight. However, the fine mode structures inside clusters are determined by signal and idler resonances. As shown in \Fig{Resonantcluster}(b), the intensity of PDC emission in the two side clusters is weaker than the central cluster. Within a single cluster, the number of resonances contributing to the PDC generation depends on the bandwidth of the resonances and the difference of FSRs.

Due to the spectral density redistribution in the cavity, PDC in the resonant waveguide is enhanced due to cavity narrowing and cluster effect. The enhancement factor $M$ is given by
\begin{eqnarray}
M  =  {{\int {d{\omega _s}} \int {d{\omega _i}} {{\left| {{f}\left( {{\omega _s},{\omega _i}} \right)} \right|}^2}} \over {\int {d{\omega _s}} \int {d{\omega _i}} {{\left| {f\left( {{\omega _s},{\omega _i}} \right)} \right|}^2}}\mathcal{A}_s \left( {\omega _s}\right)\mathcal{A}_i\left({\omega _i} \right)}
\sim  {\eta _{\rm{pp}}} \mathcal{F}_{s}\mathcal{F}_{i} N_0,
\label{eq:EnhancementM1}
\end{eqnarray}
where ${\eta _{\rm{pp}}}$ is the photon pair escape probability defined by the ratio of the number of photon pairs which leaves the the cavity to the number of pairs created inside the cavity. More details are discussed in \ref{sec:appendixlifetime}. Eq.~(\ref{eq:EnhancementM1}) tells us that the higher the finesse of cavity is, the more enhancement can be achieved \cite{OuPRL1999}. The enhancement factor is roughly proportional to the square of the finesse of the cavity.

However, for a waveguide based resonant PDC source the internal losses must be considered. (In bulk type sources, these might be neglected due to the much smaller losses in bulk crystals compared to waveguides.) The higher the finesse is, the larger the number of round-trips is. Thus, the probability that a created photon pair is lost within the cavity increases with increasing the finesse. Hence,  ${\eta _{\rm{pp}}}$  decreases with increasing the finesse by higher mirror reflectivities keeping waveguide length and losses constant. On the other hand, the spectral redistribution due to the clustering results in a $N_0$ times increased spectral density at the synchronous signal and idler resonance.

\subsection{Two-photon cross-correlation function}\label{sec:2ndcorr}
To verify the photon pairs generation and to evaluate its efficiency, the standard way is to measure the cross-correlation between two generated photons.
The two-photon cross-correlation function $G_{si}^{\left( 1,1 \right)}$ between signal and idler photons defines the shape of the coincidence events. Each resonance of the frequency comb of signal and idler can reasonably well be approximated by a Lorentzian function ${A_{s,i}}\left(
{{\omega _{s,i}}} \right){\propto }({\gamma _{s,i}}{\rm{ + }}i{\omega _{s,i}})^{-1}$, where ${{\gamma _{s,i}}}$ describes the damping constants of cavity. Assuming only a single pair of signal ${\omega _{s}}$ and idler modes ${\omega _{i}}$ are synchronously resonant with narrowband range $\delta\omega$, we get the complex JSF
\begin{eqnarray}
{f_R}\left( {\delta \omega } \right) =  {{\gamma _s} \over {{\gamma _s}{\rm{ + }}i\delta \omega}}{{\gamma _i} \over {{\gamma _i} - i\delta \omega }}.
\label{eq:SinModCavJSF02}
\end{eqnarray}

The temporal $G_{si}^{\left( 1,1 \right)}\left( \tau  \right)$ is measured as the coincidence distribution of detection time differences $\tau = t_s -t_i$ between the signal and idler photons,
\begin{eqnarray}
G_{si}^{\left( 1,1 \right)}\left( \tau  \right) = \left\langle {\hat a_s^\dag \left( t \right)\hat a_i^\dag \left( {t + \tau } \right){{\hat a}_i}\left( {t + \tau } \right){{\hat a}_s}\left( t \right)} \right\rangle = |\tilde{f}_{R} (\tau)|^2, \label{eq:G2SItheory01}
\end{eqnarray}
where $\tilde{f}_{R} (\tau)$ is the joint temporal function (JTF) defined as the inverse Fourier transform of the JSF.
Thus, the signal-idler correlation can be simplified as
\begin{eqnarray}
G_{si}^{\left( 1,1 \right)}\left( \tau  \right)  \propto
u\left( \tau  \right){e^{ - {2\gamma _s}\tau }} + u\left( { - \tau } \right){e^{{2\gamma _i}\tau }},
\label{eq:CoinSIsm01}
\end{eqnarray}
where $u\left( \tau  \right)$ is the step function. This shows that the coincidence funtion is determined by two different exponentially decaying functions, which are related to the damping of signal and idler photons, respectively. Thus, in most cases $G_{si}^{\left( 1,1 \right)}\left( \tau  \right)$ is asymmetric due to ${\gamma _{s}} \neq {\gamma _{i}}$.
The signal-idler correlation time is defined by $\tau _{c} \sim {2({\tau_s}+{\tau_i})/e}$. This means, we can identify the lifetime of signal/idler ${\tau _{s,i}}={(2\gamma _{s,i})}^{-1}$ from the signal-idler coincidence measurements. Only if ${\tau_s} ={\tau_i}$, which for instance is always the case if type I degenerate PDC process are exploited \cite{PomaricoNJP2009}, one gets a symmetric cross-correlation with with correlation time $\tau _{c} \sim {4{\tau_s}/e}$.

\subsection{Two-photon auto-correlation function}
Besides the one photon pairs generated in the PDC process, it is still possible to generate double or higher photon pairs as shown in Eq.~(\ref{eq:RPDCstate1})).
The temporal second-order auto-correlation function $G^{\left( 2 \right)}\left( \tau  \right)$ of signal and idler photons respectively, is related to the temporal coherence of the two-photon wave-packet component. The measurements of the auto-correlation is usually performed by inserting a 50:50 beam splitter into the signal (idler) arm and detecting the photons at the two output ports of the beam splitter with single photon detectors. If coincidence clicks are registered, a second (or higher) order photon pair generation must have happened. ${G^{\left( 2 \right)}}\left( \tau  \right)$ is defined by
\begin{eqnarray}
{G^{\left( 2 \right)}}\left( \tau  \right) & = & \left\langle {{{\hat a}^\dag }\left( t \right){{\hat a}^\dag }\left( {t + \tau } \right)\hat a\left( {t + \tau } \right)\hat a\left( t \right)} \right\rangle  \nonumber \\ & = & \left\langle {{{\hat a}^\dag }\left( t \right)\hat a\left( t \right)} \right\rangle \left\langle {{{\hat a}^\dag }\left( {t + \tau } \right)\hat a\left( {t + \tau } \right)} \right\rangle  \nonumber \\ & + &  \left\langle {{{\hat a}^\dag }\left( t \right)\hat a\left( {t + \tau } \right)} \right\rangle \left\langle {{{\hat a}^\dag }\left( {t + \tau } \right)\hat a\left( t \right)} \right\rangle .
\label{eq:G2ss01}
\end{eqnarray}
Thus, the normalized second-order auto-correlation function $g^{\left( 2 \right)}\left( \tau  \right)$ is given by
\begin{eqnarray}
 g^{\left( 2 \right)}\left( \tau  \right) & =&  1 + {\left| {{{\left\langle {{{\hat a}^\dag }\left( t \right)\hat a\left( {t + \tau } \right)} \right\rangle } \over {\left\langle {{{\hat a}^\dag }\left( t \right)\hat a\left( t \right)} \right\rangle }}} \right|^2} = 1 + {\left| {{g^{\left( 1 \right)}}\left( \tau  \right)} \right|^2}.
\label{eq:Norg2ss01}
\end{eqnarray}
Consider a signal cavity mode with a Lorentzian spectral distribution,
\begin{eqnarray}
{f_s}\left( {\delta \omega } \right) = {{\gamma _s} \over {{\gamma _s}{\rm{ + }}i\delta \omega}}{{\gamma _i} \over {{\gamma _i} - i\delta \omega }},
\label{eq:Cav2SignalJSF01}
\end{eqnarray}
the time-dependent correlation function is given by the inverse Fourier transform of its intensity spectrum,
\begin{eqnarray}
& &{\left\langle {{{\hat a}^\dag }\left( t \right)\hat a\left( {t + \tau } \right)} \right\rangle } \\ \nonumber
& = & \mathcal{FT}^{-1} [{f_s}\left( {\delta \omega } \right){f_s^*}\left( {\delta \omega } \right)]
=  {1 \over {4}}{\gamma _s}{\gamma _i}{e^{ - {\gamma _s}\left| \tau  \right|}} * {e^{ - {\gamma _i}\left| \tau  \right|}}\\ \nonumber
& = & \left\{ {\matrix{
   {{{{\gamma _s}{\gamma _i}} \over {\gamma _i^2 - \gamma _s^2}}\left[ {{1 \over {2}}{\gamma _s}{e^{ - {\gamma _s}\left| \tau  \right|}} - {1 \over {2}}{\gamma _i}{e^{ - {\gamma _i}\left| \tau  \right|}}} \right],} & {{\gamma _s} \ne {\gamma _i}}  \cr
   {{1 \over {4}}\gamma _s\left( {1 + {\gamma _s}\left| \tau  \right|} \right){e^{ - {\gamma _s}\left| \tau  \right|}},} & {{\gamma _s} = {\gamma _i}}  \cr} } \right..
\label{eq:AutoG202}
\end{eqnarray}
Correspondingly, Eq.~(\ref{eq:Norg2ss01}) can be deduced as
\begin{eqnarray}
&& {g^{\left( 2 \right)}}\left( \tau  \right) \\ \nonumber
& = & 1 +  \left| g_0^{\left( 1 \right)} \left( \tau  \right) \right|^2 \\ \nonumber
& = &  \left\{ {\matrix{
   {1 + {{\left| {{1 \over {{\gamma _i} - {\gamma _s}}}{e^{ - {{{\gamma _s} + {\gamma _i}} \over 2}\left| \tau  \right|}}\left( {{\gamma _i}{e^{{{{\gamma _i} - {\gamma _s}} \over 2}\left| \tau  \right|}} - {\gamma _s}{e^{ - {{{\gamma _i} - {\gamma _s}} \over 2}\left| \tau  \right|}}} \right)} \right|}^2},} & {{\gamma _s} \ne {\gamma _i}}  \cr
   {1 + {{\left| {{e^{ - {\gamma _s}\left| \tau  \right|}}\left( {1 + {\gamma _s}\left| \tau  \right|} \right)} \right|}^2},} & {{\gamma _s} = {\gamma _i}}  \cr
} } \right..
\label{eq:Norg2ss02}
\end{eqnarray}
The above equation shows that the decay of the auto-correlation is not an exponential decay any more. For our further analysis, it is useful to estimate the shape and half width of the auto-correlation peak. By using a Taylor expansion, the above complicated function can be approximately simplified to
\begin{eqnarray}
\label{eq:Norg2ap01}
{g^{\left( 2 \right)}}\left( \tau  \right) & \simeq & 1 + {\left| {{e^{ - {1 \over 2}\left( {{\gamma _s} + {\gamma _i}} \right)\left| \tau  \right|}}\left( {1 + {1 \over 2}\left( {{\gamma _s} + {\gamma _i}} \right)\left| \tau  \right|} \right)} \right|^2} \\ \nonumber
& \approx &  1 + {\left( {1 + {{\left[ {{1 \over 2}\left( {{\gamma _s} + {\gamma _i}} \right)\tau } \right]}^2}} \right)^{ - 1}}.
\end{eqnarray}
From Eq.~(\ref{eq:Norg2ap01}), we can conclude that the auto-correlation peak is symmetric, no matter which kind of PDC process it produced. This shows an intrinsic difference to the behavior of the cross-correlation function with its asymmetrically exponential decays. The decay of auto-correlation can be approximated as a Cauchy–Lorentz distribution.
We can define the auto-correlation time $T _{au}$ as the FWHM of the auto-correlation function and obtain
\begin{eqnarray}
{T _{au}} \approx {4 \over {{\gamma _s} + {\gamma _i}}}\approx  {2 \over {\ln 2}}{\tau _c}.
\label{eq:Norg2au01}
\end{eqnarray}
We find that the auto-correlation time exceeds the cross-correlation time by a factor of around 2.8. Our theoretical analysis is in good agreements with previous experimental observations \cite{HockelPRA2011, FortschNC2013}.

Another way of interpreting this auto-correlation is to consider the measurement from the multiple photon pair generation point of view. If two single photon detecdtors in the signal (or idler) arm register a coincidence click, the second (or higher) order part of the PDC state (second term in Eq.~(\ref{eq:RPDCstate1})) is probed. In contrast, the cross-correlation measurement provides mainly information about the first order part. The auto-correlation time $T _{au}$ is the correlation time between the two-photon components of signal (or idler) photon components, while the cross-correlation time $\tau _{c}$ is the correlation time between signal and idler photons. In the temporal domain, the auto-correlation time $T _{au}$ is larger than correlation time $\tau _{c}$. Correspondingly, two photon components have narrower frequency bandwidth (related to ${f \left( {{\omega _s}{\rm{,}}{\omega _i}} \right)}{f \left( {{\omega _s'}{\rm{,}}{\omega _i'}} \right)}$) than the one photon pair component (related to ${f \left( {{\omega _s}{\rm{,}}{\omega _i}} \right)}$).

\subsection{Heralded correlation function}
To confirm the nonclassical property of generated narrowband photons, it is important to characterize heralded photons.
The conditional auto-correlation function is defined by \cite{BettelliPRA2010}
\begin{eqnarray}
g_{\rm{heralded}}^{\left( 2 \right)}\left( {{t_{s1}},{t_{s2}}\left| {{t_i}} \right.} \right) = {{P_{ssi}\left( {{t_{s1}},{t_{s2}},{t_i}} \right)} \over R ^3{g_{si}^{\left( 1,1 \right)}\left( {{t_{s1}} - {t_i}} \right)g_{si}^{\left( 1,1 \right)}\left( {{t_{s2}} - {t_i}} \right)}},
\label{eq:ConG2ssi01}
\end{eqnarray}
where ${P_{ssi}\left( {{t_{s1}},{t_{s1}},{t_i}} \right)}$ is the triple coincidence rate, and $R$ is the pair generation rate. After simplification, the function of interest, $g_{\rm{heralded}}^{\left( 2 \right)}\left( \tau  \right)= g_{\rm{heralded}}^{\left( 2 \right)}\left( {{t_i},{t_i} + \tau \left| {{t_i}} \right.} \right)$ can be represented as
\begin{eqnarray}
g_{\rm{heralded}}^{\left( 2 \right)}\left( \tau  \right) \approx 1 - \rm{exp}{\left[ {- 2 { {\left(\frac{ {{\gamma _s} {\gamma _i}} }{ {{\gamma _s} + {\gamma _i}} } \tau \right)} ^2}} \right]},
\label{eq:ConG2s02}
\end{eqnarray}
which means the heralded-correlation can be approximated by a Gaussian distribution.
$T _{he}$ is the heralded time between one (signal or idler) photon component,
\begin{eqnarray}
{T_{he}} \approx {2 \over {{\gamma _s} + {\gamma _i}}}= {1 \over 2}{T _{au}}.
\label{eq:ConGg2the01}
\end{eqnarray}

\section{Waveguide resonator fabrication}\label{Fabrication}

\begin{figure}
\begin{center}
\includegraphics*[width=0.9\textwidth]{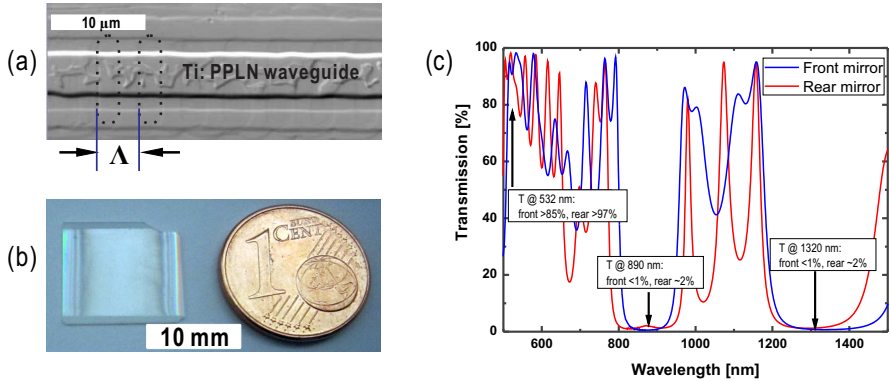}
\end{center}
\caption{Integrated narrowband photon pair source composed of a Ti:PPLN waveguide with $\Lambda=4.44~\mu$m poling period and dielectric mirrors deposited on the waveguide end-faces.
(a)Top view onto the Ti:PPLN waveguide with a zoom to the domain structures. The black dotted lines denote the inverted domains. (b) The photograph shows the actual size of the miniaturized device compared to a one cent euro coin (16.25 mm diameter).
(c) Spectral characteristics of the cavity mirrors deposited on the waveguide endface. The measurements have been performed on witness substrates coated simultaneously with the waveguide sample.}
\label{ResonantWG}
\end{figure}

To study experimentally such a resonant PDC source, a PPLN waveguide with dielectric coatings at its end faces was fabricated.
It was designed to provide type II phase-matching for a PDC process with a TE-polarized pump at 532~nm for the generation of photon pairs at 890 nm (TE-polarized) and 1320 nm (TM-polarized).

\subsection{Ti:PPLN waveguide}
The waveguide was fabricated by an indiffusion of a lithographically patterned,  81 nm thick, 5 $\mu$m wide Ti-stripe into the Z-cut LiNbO$_3$ substrate. The diffusion parameters were designed to provide single mode guiding at the TM-polarized idler mode. At signal and pump wavelength, however, the waveguide supports several guided modes.

Subsequent to the Ti-indiffusion, the periodic poling was performed. To obtain first order phase-matching for the selected wavelength combination a short poling period of only $\Lambda\approx4.44~\mu$m is required. This could be achieved using a field-assisted domain inversion of the lithographically defined poling pattern.
In the final fabrication process the end-faces were polished. Special care was taken to ensure flatness and parallelism of the waveguide end-faces to avoid excess loss of the resonating waves at the multiple reflections at these interfaces.

\subsection{Resonator}
Based on the theoretical design rules  discussed in the previous section and \ref{sec:appendixfinesse}, the resonator finesse should exceed 20 to provide the spectral narrowing. The detailed structure of the integrated waveguide chip and experimental setup to study this resonant source are shown in \Fig{ResonantWG}. The source consists of a 12.3~mm long Ti-indiffused waveguide in Z-cut PPLN and dielectric mirrors deposited on the waveguide end-faces.

To implement the resonant source, dielectric layers composed of alternating layers of SiO$_2$ and TiO$_2$ were deposited on the end-faces of the waveguide. In practice, a stack with 17 layers deposited as front mirror provides a reflectivity of $R_{s,i}\approx99\%$ for signal and idler wavelengths and the rear mirror consisting of 17 layers has the targeted $R_{s,i}\approx98\%$ for both. The reflectivities of front and rear mirrors at 532~nm are around 15 \% and 13 \%, respectively, to enable efficient incoupling of pump and to prevent triple resonance effects.

After mirror deposition, the resonator was characterized by measuring signal and idler decay times. As we discuss in \ref{sec:appendixlifetime}, we obtained a finesse of  $\mathcal{F}\sim 100$ and $\mathcal{F}\sim 80$ for the signal and idler wavelengths, respectively.

With the asymmetry of the mirror reflectivities, the ratio of outcoupled signal (idler) photons to lost photons before escaping this cavity is about $\eta_s \approx 0.28$ ($\eta_i \approx 0.18$), respectively. Thus, the overall photon pair escape probability is given as ${\eta _{pp}} = {\eta _s}{\eta _i}$, which means about 5\% of the generated photon pairs leave the cavity as couples at the desired output mirror. Therefore, the cavity-enhanced source provides high heralding efficiency.

The spatial profiles of signal and idler are determined by the waveguide properties. Both are the fundamental modes of the waveguide. Thus, the profiles are well matched for fiber network applications.

\section{Experimental setup}\label{Setup}

\begin{figure}[tbp]
\includegraphics*[width=0.9\textwidth]{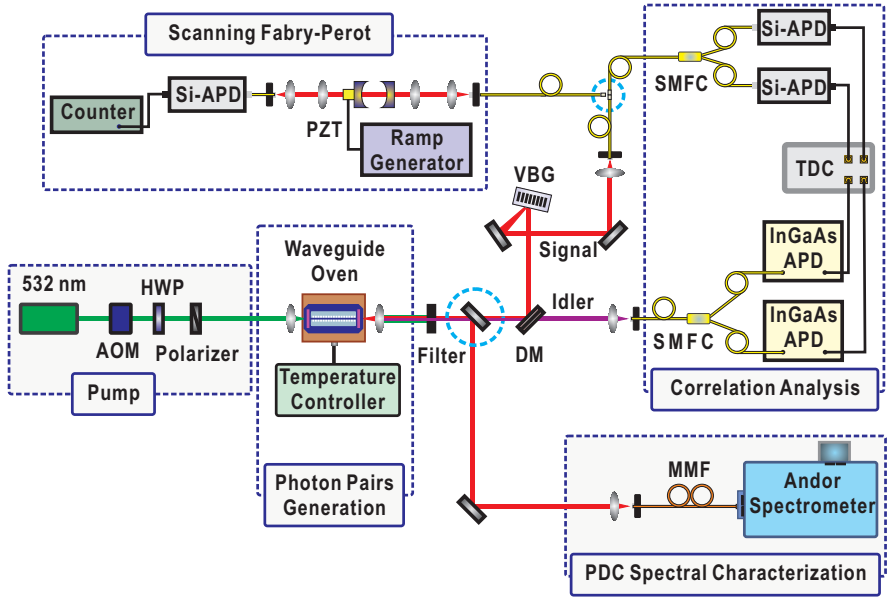}
\caption{ Experimental setup for the generation and characterization of
the photon pairs,
combining PDC characterization and coincidence measurements.
The cyan dashed circle means the optical beam paths are alternative.
AOM: acousto-optical modulator; HWP: half-wave plate; DM: dichroic mirror;
SMFC: single-mode fiber coupler; MMF: multi-mode fiber; VBG: volume Bragg grating;
PZT: piezoelement; APD: avalanche photodiode; TDC: time-to-digital converter.}
\label{QuRePSetup}
\end{figure}

The experimental setup to study this resonant source is shown in \Fig{QuRePSetup}.
The sample is pumped with a laser at 532 nm
with a specified bandwidth of less than 1 MHz.  To mitigate photorefraction,
pump pulses with a typical
length of about 200 ns and a repetition rate of about 100 kHz are extracted from
the cw-source by using an acousto-optical modulator (AOM). A half wave plate (HWP)
together with a polarizer enable pump peak power tuning from 0.1~mW to 10~mW.
The sample is heated to temperatures around 160~$^{\circ}$C to
obtain quasi-phase-matching for the desired wavelength
combination and to prevent luminescence and deterioration due to photorefraction. During the
measurements the sample temperature is stabilized to about $\pm$1 mK. Coarse wavelength tuning can be accomplished by changing the poling period, while
fine tuning of the resonance frequency is accomplished by varying the optical path length
in the waveguide, for instance by temperature tuning.

To characterize the generated PDC spectrum, the signal output from the waveguide is coupled to a spectrometer system (Andor Shamrock 303i with iKon-M 934 CCD camera)  with a resolution of
about 0.15~nm.

This measurement method  provides a coarse overview over the spectral structure, but the resolution is still not sufficient to study details of the spectra within a single cluster. Thus, a volume Bragg grating (VBG, OptiGrate 900) with a spectral bandwidth
of 0.17 nm is inserted into the signal beam path as shown in \Fig{QuRePSetup} to act as bandpass filter to select a single cluster. The filtered light is routed via a single mode fiber to a scanning
confocal Fabry-P\'{e}rot resonator with 15 GHz FSR and a finesse of about 20. Its transmission is analyzed using
an avalanche photodiode (APD, Perkin Elmer SPCM-AQR-14)
to record the signal photons transmitted from the Fabry-P\'{e}rot  which is  tuned by applying a voltage ramp to the piezo-driven mirror mount.

To study correlations between photon pairs, coincidences between signal and idler were characterized by measuring the arrival times of the respective photons with the set-up shown in Fig.~\ref{QuRePSetup}.
The cross-correlation could be determined measuring coincidences between signal and idler photons by spectrally splitting the photon pair and routing them to separate single photon detectors (Perkin Elmer single photon APD for the signal and idQuantique InGaAs APD model id201 for the idler photons). The signal-signal autocorrelation were performed by splitting  the signal radiation behind the VBG (i.e.\ a single cluster was selected) using a 50:50 single mode fiber coupler (SMFC) and routing the two SMFC's output ports to individual silicon APDs. Similarly, there are two InGaAs APDs to catch unfiltered idler photons behind a SMFC.

\section{Results and discussion}\label{Results}

\begin{figure}
\begin{center}
\includegraphics*[width=0.9\textwidth]{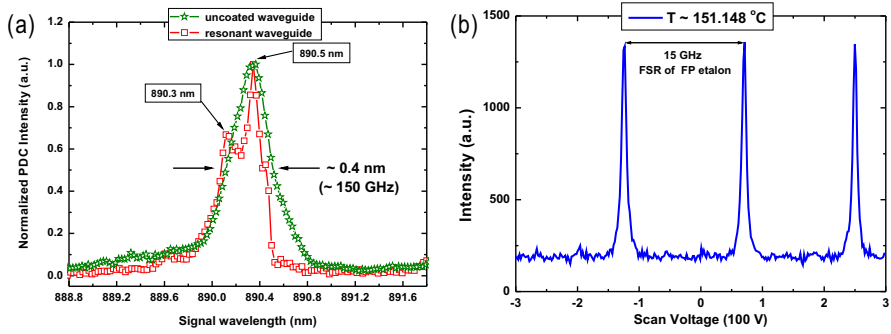}
\end{center}
\caption{\label{NarrowPDC} (a) Enhanced PDC spectra.
PDC spectra measured using the uncoated waveguide (green stars)
and the resonant waveguide (red squares).
The background counts are due to the dark counts of the camera.
(b) Signal fine spectra at single-photon level in one cluster.
 Signal spectra recorded with a
confocal scanning Fabry-P\'{e}rot with 15 GHz FSR.
The two spectra vary in temperature by 6~mK around 161.57~$^{\circ}$C.
}
\end{figure}

\subsection{PDC clustering and cavity-enhanced spectra}

PDC spectra of one waveguide recorded prior and after mirror deposition
are shown in \Fig{NarrowPDC}. The spectrum of the uncoated waveguide shows a main peak with
a bandwidth (FWHM) of 0.4~nm ($\approx$ 157~GHz), as predicted for the interaction length. The spectra
of the same waveguide with dielectric mirrors, however, shows a pronounced sub-structure.
Although the resolution of the spectrometer ($\approx 0.15$~nm) is
too coarse to reveal details of the spectra, one can already derive that the pair generation occurs within clusters with a cluster separation of about 0.24~nm ($\approx$ 90~GHz). This separation is determined by the beating
of the FSRs of signal and idler, which theoretically are FSR$_s\sim5.2$~GHz and FSR$_i\sim5.5$~GHz.
Generally, there is one dominant cluster in the center of the normalized phase-matching curve, and another two symmetric clusters at the side wings of the envelope, as simulated in \Fig{Resonantcluster}(b).

A comparison of the spectral measurements of the coated and uncoated sample proves already the enhancement of PDC generation in the cavity: If the cavity would only act as a comb filter, only a small fraction of the generated PDC spectrum would 'transmit' through this filter. In this case we would have seen only a weak signal for the resonant PDC because the measured spectra are always convolutions of the real spectra with the spectral resolution, i.e. a window function of about 60 GHz ($\approx 0.15$~nm) width. In the experiments, however, we observed in both cases spectra with similar intensity levels. For the measurements shown in \Fig{NarrowPDC} (a), all of the parameters, apart from adapting the temperature, have been kept constant. In particular, the pump power of about 1~mW in front of the incoupling lens has not been changed. Thus, we can conclude that due to the cavity there is a spectral density redistribution resulting in an enhanced PDC generation at the resonances.

To study the spectral structure within a single cluster we performed measurements with the scanning Fabry-P\'{e}rot resonator. Measured single-photon level PDC spectra with $\sim$ 700 MHz resolution are shown in \Fig{NarrowPDC}(b). They reveal
the modal structure within a single cluster.  The spectrum consists of only one longitudinal mode. The scanning range of
15~GHz covers around 3 FSR of the signal resonator. Please note that the measured linewidth of each longitudinal mode of about 700 MHz is mainly determined by the resolution of the scanning Fabry-P\'{e}rot, but it is not the natural bandwidth of the generated PDC photons.

\begin{figure}
\begin{center}
\includegraphics*[width=0.7\textwidth]{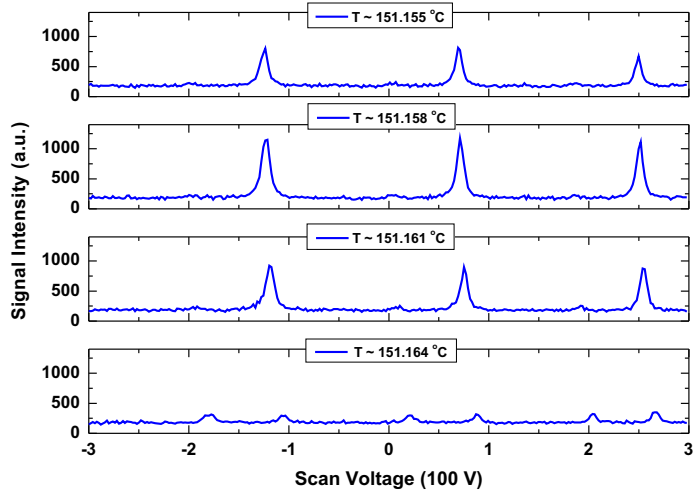}
\end{center}
\caption{\label{NarrowPDCTscan}
Signal fine spectra at single-photon level in one cluster recorded at different temperatures around 151.16~$^{\circ}$C. Multiple peaks in the scans separated by around 200 V correspond to subsequential resonances of the analyzing Fabry-P\'{e}rot cavity.}
\end{figure}

The single-photon level PDC spectra with GHz resolution indicates that we are able to suppress the adjacent 'satellite' modes to single mode operation by finely tuning the temperature -- as shown in \Fig{NarrowPDCTscan}. A slight shift of the temperature
leads to the spectrum composed of always one predominant mode. Only at certain temperature, weak two mode operation happens.

\subsection{Cross-correlation}
The temporal second-order signal-idler cross-correlation function \cite{Glauber1963} is measured as the coincidence counts between the signal and idler photons as function of the arrival time difference (more details were already discussed in \ref{sec:2ndcorr}).

\begin{figure}
\begin{center}
\includegraphics*[width=0.9\textwidth]{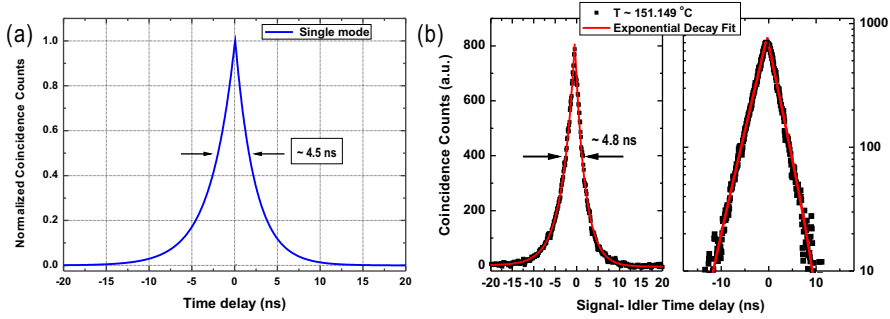}
\end{center}
\caption{\label{NarrowG2} (a) Simulated coincidences between signal and idler photon generated from
the double cavity. The blue curve is corresponding to the coincidence between signal and idler, which are ideally single mode.
(b) Signal-idler coincidences.
Measured coincidences of photon pairs
as a function of arrival time difference between
the signal and idler photons using linear (left) and logarithmic (right) scaling with exponential fits.
}
\end{figure}

To compare the experimental result with the theoretically expected behaviour we first simulated the cross-correlation function for the used resonator parameters assuming a truly single mode PDC generation as shown in \Fig{NarrowG2}(a).
The presence of the cavity implies that signal and idler photons may be emitted at distinct times, corresponding to a different number of round-trips within the cavity. Thus, the shape of the coincidence curve should be determined by exponential functions.
For the multi-mode case the distribution of time of emission differences between the signal and idler modes should show well defined peaks with the peak separation corresponding to the cavity round--trip time. However, in our experiment we could not resolve such oscillations due to the finite temporal resolution of our detection system of about 500 ps. The convoluted results concerning the resolution of detector system is a smooth exponential decay curve as well. Taking into account this limitation again by a convolution of the simulated curves,  we find that the bandwidth of the coincidence peak is independent of the number of modes, and depends only on the bandwidth of each modes.

In the left diagram of \Fig{NarrowG2}(b), a result of such a coincidence measurement is shown. It reveals a correlation time (FWHM of the coincidence peak) of about 4.8~ns. This is significantly broader than the corresponding results obtained with non-resonant samples showing a width of about 0.5~ns, which is determined by the finite resolution of our measurement system. The correlation time is inversely proportional to the bandwidth of the down-conversion fields.
From the measured correlation time $\tau_{c}$ of 4.8~ns, according to $\tau_{c}=1/{\pi \Delta\nu}$, where $\Delta\nu$ is the bandwidth of the down-converted photons, a spectral bandwidth of about 66~MHz can be deduced. This is in good accordance with the theoretically predicted width calculated for the given cavity parameters, as shown in \Fig{NarrowG2}(a).
In the right diagram of \Fig{NarrowG2}(b) the measurement result is redrawn using a logarithmic scaling together with exponential fits for the rising and falling parts. The slight asymmetry reflects the different finesses of signal and idler resulting in slightly different leakage times out of the resonator. The different slopes of decay lines in logarithmic scaling give us the decay times of resonators.

Based on the coincidence data and \ref{sec:appendixlifetime},  the different lifetimes of signal and idler leaving out from the resonant waveguide are $\tau_{s} \sim 3$~ns, and $\tau_{i} \sim 2.4$~ns, respectively. Thus, the loss inside the waveguide cavity can be determined to be as low as $\alpha_s \approx 0.016$~dB/cm and $\alpha_i \approx 0.022$~dB/cm for signal and idler, respectively, as well as  cavity finesses of $\mathcal{F}_s\simeq100$ and $\mathcal{F}_i\simeq80$.

\subsection{Brightness}
The coincidence measurements can also be evaluated to determine the efficiency of the PDC generation process.
From the ratio of the coincidence to single counts the generated photon pair rate can be determined. We found that this rate is almost equal for non-resonant and resonant waveguides. For our source we determined a normalized generation rate (inside the resonator) of about $50\times10^6~$pairs/(s$\cdot$mW). Assuming these are distributed over three inequivalent clusters with different weights, we can estimate that about 90\% of the generated photon pairs are within the most dominant longitudinal mode with 66~MHz bandwidth. Taking into account the photon pair escape probability of  5~\%, the spectral brightness can be estimated to be  $B = 3\times10^4~$pairs/(s$\cdot$mW$\cdot$MHz).

\begin{figure}[tbp]
\includegraphics*[width=0.9\textwidth]{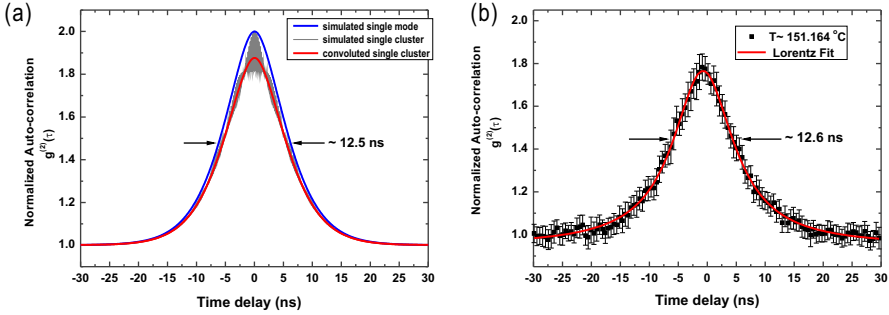}
\caption{(a) Simulated temporal auto-correlations of signal photon. The blue envelope is for single mode case; the gray and the red are for multi-mode signal within one cluster without and with limited resolution, respectively.
(b)Experimental signal-signal auto-correlation.
Measured signal-signal auto-correlation as function of arrival time difference between
two signal photons within one cluster.}
\label{NarrowFFTtime}
\end{figure}

\subsection{Auto-correlation}

An alternative method to characterize the spectral properties of the source is the analysis of the auto-correlation Glauber function $g^{(2)}(\tau)$. The number of cavity modes $N$ can be estimated from the measured normalized auto-correlation function at zero time delay according to  $g_m^{(2)}(0)=1 + 1/ N$ (more details are given in \ref{sec:appendixmultimode}).

\Fig{NarrowFFTtime}(a) provides the simulated and convoluted auto-correlation results for signal photons from the resonant waveguide. In the ideal case of a pure single cavity mode field, the second-order auto-correlation is close to 2 at zero time delay. For a cluster obtained from this resonant waveguide, which consists of several inequivalent modes (as shown in \Fig{Resonantcluster}(b) and \Fig{NarrowPDC}(b)), the temporal auto-correlation behavior should reveal multi-mode interference as well. However, we can not resolve the interference fringes due to the resolution of the detection system. As a results, only one single convoluted peak with lower $g_m^{(2)}(0)$ could be observed in the experiment.

From the measured value $g_m^{(2)}(0)\approx 1.85$ shown in \Fig{NarrowFFTtime}(b), an effective cavity mode number of $N\simeq1.2$ can be estimated. This is in reasonably good qualitative agreement with the simulated auto-correlation function shown in \Fig{NarrowFFTtime}(a) and the measured spectra shown in \Fig{NarrowPDC}, where we observed only one mode within a single cluster. The signal auto-correlation time ($\sim$ 12.6~ns) is about 2.8 times broader than signal-idler correlation time ($\sim$ 4.8~ns), which is in good agreement with the theoretical analysis (Eq.~(\ref{eq:Norg2au01})). The Lorentz fit to the experimental curve also coincides with our theory model.
Moreover, it indicates that shaping single photon wavepackets by using amplitude and phase modulators is possible \cite{HarrisPRL2008}, because the signal auto-correlation time is much longer than the time jitters from detectors and electro-optic modulators.

\subsection{Heralded correlation}
\begin{figure}[tbp]
\includegraphics*[width=0.7\textwidth]{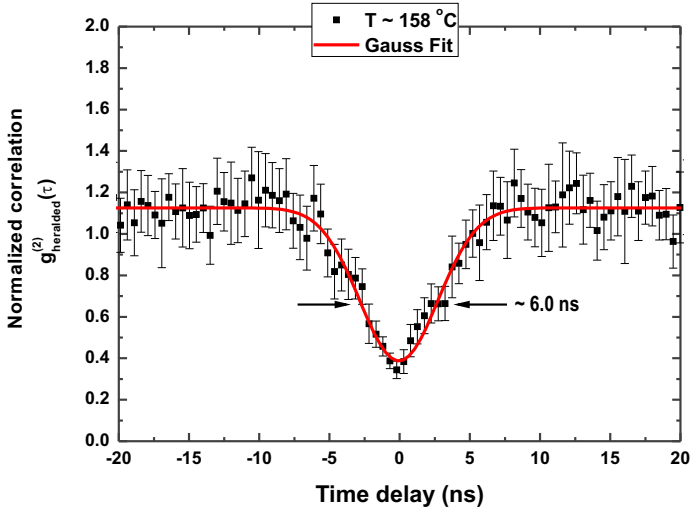}
\caption{
Measured heralded correlation $g^{(2)}_{\rm{heralded}}(\tau)$ and its Gauss fit as function of arrival time difference between
two signal photons within the whole phase-matching range operated around 158~$^{\circ}$C.}
\label{Narrowg2he}
\end{figure}

To understand the intrinsic nature of the source, it is important to investigate the heralding performance for this resonant waveguide. The heralding efficiency would suffer from the asymmetric reflectivities, because the photon pairs escaping probability is impacted by the reflectivities of both dielectric. However, the signal and idler long wave-packets generated from the resonant waveguide could help to resolve the wave-packet of single signal photons, due to the broadened temporal correlation length. The result of normalized heralded second-order correlation function is shown in \Fig{Narrowg2he}. The $g_{\rm{heralded}}^{(2)}(0) < 0.3$ at high pump power (10 mW) confirms a genuine sign of non-classicality. By using low power pumping ($\sim$ 1 mW), the minimum droped to $g_{\rm{heralded}}^{(2)}(0) < 0.02$.

Comparing heralded correlation with auto-correlation, it is found that they have different distributions and bandwidths. The heralded one($\sim$ 6.0~ns) is roughly half of the two-photon auto-correlated one ($\sim$ 12.6~ns). The higher-photon components contributions to heralded correlation measurements mainly influence the minimum value, since the probability to generate single photon pairs is absolutely higher than the probability to generate double or even triple photon pairs.

\section{Conclusion}\label{sec:conclusion}
In summary, we have experimentally demonstrated a compact single-longitudinal-mode two-color narrowband photon pair source in a doubly resonant Ti:PPLN waveguide. This elaborate source with its compact design emerges multiple strengths, such as narrow bandwidth, single longitudinal mode, high brightness, addressing atom and fiber compatibility. Our work shows that clustering can be exploited to restrict PDC generation to a single longitudinal mode. We have also observed several novel experimental phenomena behind the cavity-enhanced PDC, like cross-correlation time is smaller than auto-correlation time, and two-photon components have longer wave-packets than single ones.

Although the measured bandwidth of about 60~MHz is still about one order of magnitude larger than the smallest bandwidth demonstrated with bulk optical versions \cite{FeketePRL2013}, our integrated source outperforms the bulk optic realizations in many key features (see also the comparison in Tab.~\ref{tb:RefResult01}). In particular, the rugged design, pure wave-packet and large brightness of about $3\times10^4~$pairs/(s$\cdot$mW$\cdot$MHz) are excellent properties of our device making it a versatile source for various quantum applications,
for instance to address a Nd-based QM with a bandwidths of around 120~MHz \cite{GisinN2011} or Raman QM with a bandwidths of a few GHz \cite{Michelberger2014} and quantum networks \cite{ThewOE2014}. In general, our source has a great potential for various QCIP applications due to pure state, excellent temporal behavior and good compatibility with fiber based systems. Additionally, our device paves the way to arbitrarily shaped single photons in the temporal
domain by using integrated devices. Future activities can focus on the development of pure single mode sources using pulsed light and pulse shaping in resonant waveguide \cite{KalachevPRA2010}. Combining clustering with double-pass pumping to suppress PDC generation to a (filter-free) single cluster operation \cite{ChuuAPL2012} and monolithic integration of an electro-optic modulator for fine tuning are further attractive goals.

\section{Acknowledgments}

The authors thank Andreas Christ, Regina Kruse, Malte Avenhaus and Michael Stefszky for useful discussions and helpful comments. This work was supported by the European Union through the QuReP project (no. 247743).

\begin{appendix}

\section{Determining the finesse}\label{sec:appendixfinesse}

In order to ensure that only a single combination of signal and idler modes is in resonance within a cluster, the difference of the free spectral range (FSR) for signal and idler must be larger than the spectral width of the single resonances. Assuming that ${FSR}_s > {FSR}_i$ , this leads to the following condition:
\begin{eqnarray}
{FSR}_s - \frac{{\Delta {\nu _s}}}{2} & > & {FSR}_i + \frac{{\Delta {\nu _i}}}{2},
\label{eq:CluasterSinglemode1}
\end{eqnarray}
where $\Delta{\nu _{s,i}}$ are the resonance bandwidth for signal and idler, respectively.
That is, the difference between different FSRs is required to be
larger than the sum of two half-widths from signal and idler. As $\Delta{\nu _{s,i}} = {FSR}_{s,i}/{\mathcal{F}_{s,i}}$, it follows
\begin{eqnarray}
\left| {{FSR}_s - {FSR}_i} \right| & > & \frac{1}{2}
\left| {\frac{{{FSR}_s}}{{{\mathcal{F}_s}}} + \frac{{{FSR}_i}}{{{\mathcal{F}_i}}}} \right|.
\label{eq:CluasterSinglemode1}
\end{eqnarray}

From this expression, we can determine the finesse to have only a single mode in the
cluster by suppressing adjacent longitudinal modes.
\begin{eqnarray}
\frac{{2{\mathcal{F}_s}{\mathcal{F}_i}}}{{{n_{gi}}{\mathcal{F}_i} + {n_{gs}}{\mathcal{F}_s}}}
&\ge& \frac{1}{{\left| {{n_{gi}} - {n_{gs}}} \right|}}.
\label{eq:CluasterFinesse1}
\end{eqnarray}
If we assume that the cavity has the same finesse for signal and idler, i.e.
$\mathcal{F}_s =\mathcal{F}_i =\mathcal{F}_i$, the finesse is given by
\begin{eqnarray}
\mathcal{F}&\ge& \frac{1}{2}\left| {\frac{{{n_{gi}} + {n_{gs}}}}{{{n_{gi}} - {n_{gs}}}}} \right|.
\label{eq:CluasterFinesse2}
\end{eqnarray}

\section{Photon pair generation rate and lifetime}\label{sec:appendixlifetime}

In order to design a single-mode source with an integrated waveguide resonator
with a precise value of the finesse, one has to optimize the reflectivity of
rear mirror and the length of waveguide to take into consideration the losses inside the waveguide.

First, let us pay attention on the probability that a photon remains
after a round-trip in the resonator, ${P_{\rm{remain}}} = {R_1}{R_2}{e^{ - 2\alpha L}}$, where $R_{1}$ and $R_{2}$ are the reflectivities of front and rear mirrors, respectively; and $\alpha$ is internal waveguide loss coefficient. This means,
the probability that a photon is 'lost' in a round-trip is given as:
\begin{eqnarray}
{P_{\rm{lost}}} = 1- {R_1}{R_2}{e^{ - 2\alpha L}}.
\label{eq:PhotonProbability1}
\end{eqnarray}
Obviously, the 'losses' of such a resonator concept are larger compared to the losses of a
single pass device. Due to multiple round-trips in the lossy waveguide, the probability that a
photon is lost, is increased. As a result, the probability that a photon generated by the PDC process is
coupled out of the resonator from the rear mirror $R_2$ can be estimated as
\begin{eqnarray}
\eta _{s,i}  = \frac{{1 - {R_2}}}{{1 - {R_1}{R_2}{e^{ - 2 \alpha_{s,i} L}}}}.
\label{eq:PhotonUseful1}
\end{eqnarray}

For a pair of photons generated by a PDC process, the photon pair escape probability ${\eta _{\rm{pp}}}$,
i.e. the probability that a generated pair leaves the resonator at the 'useful' port (not necessarily in the same round-trip),
is given as
\begin{eqnarray}
{\eta _{\rm{pp}}} = {\eta _s}{\eta _i} = \frac{{1 - {R_{2s}}}}{{1 - {R_{1s}}{R_{2s}}{e^{ - 2{\alpha _s}L}}}}\frac{{1 - {R_{2i}}}}{{1 - {R_{1i}}{R_{2i}}{e^{ - 2{\alpha _i}L}}}}.
\label{eq:PhotonPairProbability1}
\end{eqnarray}
Therefore, the efficiency of such resonant photon pair sources is
\begin{eqnarray}
{\eta _{\rm{source}}} = {\eta _{\rm{int}}} \times {\eta _{\rm{pp}}},
\label{eq:Sourceefficiency1}
\end{eqnarray}
where $\eta_{\rm{int}}$ is the internal generation efficiency in the waveguide.

Besides the source efficiency, we are also interested in the lifetime $\tau_{s,i}$ of signal/idler photons in waveguide resonator,
\begin{eqnarray}
\tau_{s,i} = \frac{{{n _{gs,i} L}}}{{c \ \rm{ln}(\sqrt{{R_{1s,i}}{R_{2s,i}}}{e^{ - \alpha_{s,i} L}})}}.
\label{eq:PhotonLifetime1}
\end{eqnarray}
Interestingly, the cross-correlation study between signal and idler allows the determination of the cavity performance at signal and idler simultaneously. From the different decay times of the cross-correlation measurements, we can deduce the loss inside the waveguide cavity and the finesse as well at the same time, when the mirror reflectivities are known.
\begin{eqnarray}
\mathcal{F}_{s,i} = \frac{\pi}{1- e ^{-\frac{n _{gs,i} L}{c\tau_{s,i}}}}.
\label{eq:LifeFinesse1}
\end{eqnarray}

\section{Multi-mode correlation function}\label{sec:appendixmultimode}
Assuming several pairs of signal ${\omega _{sn}}$ and idler modes ${\omega _{in}}$ with narrowband range $\delta\omega$  generated within the broadband phase-matching envelope. Thus, we have a complex JSF
\begin{eqnarray}
{f_R}\left( {\delta \omega } \right) = \sum\nolimits_n s_n {{\gamma _s} \over {{\gamma _s}{\rm{ + }}i(\delta \omega+ {\omega _{sn}})}}{{\gamma _i} \over {{\gamma _i} - i(\delta \omega- {\omega _{in}}) }},
\label{eq:CavJSF02}
\end{eqnarray}
where $s_n$ are different phase-matching weights.

In the multi-mode case, the time-dependent signal-idler correlation function can be generalized as
\begin{eqnarray}
\label{eq:CoinSImm01}
&& G_{si}^{\left( 2 \right)}\left( \tau  \right)  \\ \nonumber & \propto &
u\left( \tau  \right){e^{ - 2{\gamma _s}\tau }}{\left| {\sum\nolimits_n s_n {{e^{ - i{\omega _{sn}}\tau}}} } \right|^2} + u\left( { - \tau } \right){e^{2{\gamma _i}\tau }}{\left| {\sum\nolimits_n s_n{{e^{i{\omega _{in}}\tau}}} } \right|^2}.
\end{eqnarray}
It is found that the coincidence function includes interference terms resulting in a multi-mode beating under the exponential decay envelope.

If the difference between FSRs is negligible, there are a lot of cavity modes inside one cluster. First, we consider that $N$ signal (idler) resonances occur at distinct central frequencies ${\omega _{sn}}$ (${\omega _{in}}$) with equal spacing $\Delta_{Fs} = 2 \pi \rm{FSR}_s $ ($\Delta_{Fi}= 2 \pi \rm{FSR}_i$). Thus, we rewrite Eq.~(\ref{eq:CoinSImm01}) as
\begin{eqnarray}
G_{si}^{\left( 2 \right)}\left( \tau  \right)  & \propto &
u\left( \tau  \right){e^{ - 2{\gamma _s}\tau }}\left[ {\sum\limits_{n = 1}^{N }s_n^2 + \sum\limits_{n = 1}^{N - 1} \sum\limits_{j = 1}^{N-n}2s_j s_{j+n} {\cos \left( {n{\Delta_{Fs}}\tau } \right)} } \right] \\ \nonumber
& + &  u\left( { - \tau } \right){e^{2{\gamma _i}\tau }}\left[ {\sum\limits_{n = 1}^{N }s_n^2 + \sum\limits_{n = 1}^{N - 1} \sum\limits_{j = 1}^{N-n}2s_j s_{j+n}{\cos \left( {n{\Delta_{Fi}}\tau } \right)} } \right].
\label{eq:CoinSImm02}
\end{eqnarray}
This reveals that the two-photon coincidence from a multi-mode cluster is not only determined by two exponential cavity decay times. Additionally, the superposition of different mode beatings contributes from $n{\Delta_{Fs,i}}$. From the above equation, it can be seen that the coincidence distribution consists of two combs of peaks with equal separation, which corresponds to different round-trip times.
Then, if the difference between FSRs is large enough, only one mode in each cluster appears.
Considering the cluster effect, we rewrite Eq.~(\ref{eq:CoinSImm01}) as
\begin{eqnarray}
G_{si}^{\left( 2 \right)}\left( \tau  \right)  & \propto &
[u\left( \tau  \right){e^{ - 2{\gamma _s}\tau }} +  u\left( { - \tau } \right){e^{2{\gamma _i}\tau }}] \\ \nonumber
&\times&\left[ { C_1 + 2 C_2 {\cos \left( 2 \pi{{\Delta \nu_{\textrm{cluster}}}\tau } \right)} + 2 C_3 {\cos \left( 4 \pi {{\Delta \nu_{\textrm{cluster}}}\tau } \right)}} \right],
\label{eq:CoinSIcm03}
\end{eqnarray}
where $C_i$ are normalized constants from corresponding phase-matching weights. Both, information of cavity mode-beating and cluster modes interference, are included in the two-photon cross-correlation. If we have a stable PDC source with two or three single-mode clusters and a precise detection system, we would also obtain the cluster separation ${\Delta \nu_{\textrm{cluster}}}$ from such coincidence measurements.

Generally, for multi cavity modes and three clusters case, the coincidence relationship is more complicated, because mode beating and cluster beating both are involved. If the detection system with perfect resolution could resolve the emission time from different round-trips, we could characterize the cavity performance and distinguish the modes performance, like FSR and cluster separation. However,the description of realistic measurements must take into account the limited temporal resolution
of the experimental system.  Actually, the time jitter of detector systems at least covers several periods of mode beating and even more periods of cluster beating. Thus, we observed in the coincidence experiment the convolution between the real function and a window function, which is determined by the temporal resolution.

Consider a multi-mode beam with an equally spectral superposition of several Lorentzian distributions,
\begin{eqnarray}
{f}\left( {\delta \omega } \right) = \sum\nolimits_n {{\gamma _s} \over {{\gamma _s}{\rm{ + }}i(\delta \omega+ {\omega _{sn}})}}{{\gamma _i} \over {{\gamma _i} - i(\delta \omega + {\omega _{sn}}) }},
\label{eq:Cav2SignalJSF03}
\end{eqnarray}
The auto-correlation function gives
\begin{eqnarray}
\label{eq:Norg2ss03}
& & {g^{\left( 2 \right)}}\left( \tau  \right) \\ \nonumber
& = & 1 + {1 \over N}{\left| {{g_0}\left( \tau  \right)} \right|^2} + {1 \over {{N^2}}}\sum\limits_{n = 1}^{N - 1} {2\left( {N - n} \right)} \cos \left[ {n{\Delta _{F}}\tau } \right]{\left| {{g_0}\left( \tau  \right)} \right|^2},
\end{eqnarray}
where $\Delta _{F}$ is the angular FSR and $N$ is the number of modes. When $N=1$, the third interference term vanishes. In this case, Eq.~(\ref{eq:Norg2ss03}) is the same as Eq.~(\ref{eq:Norg2ss02}).

\begin{figure}[tbp]
\includegraphics*[width=0.7\textwidth]{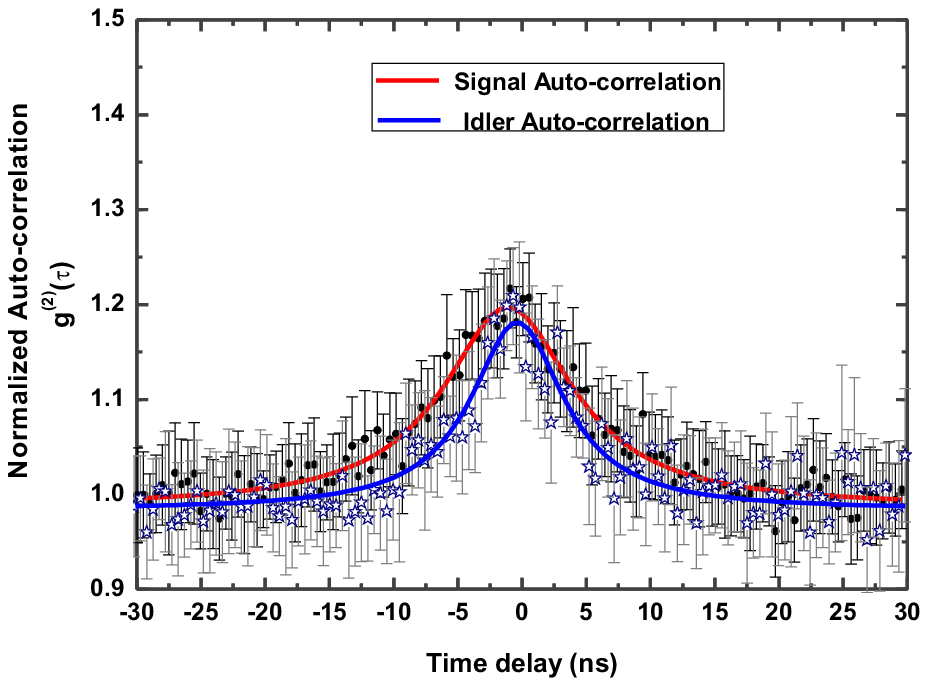}
\caption{Experimental signal-signal auto-correlation and idler-idler auto-correlation without any cluster filters.
Measured $g^{(2)}_{ss}(\tau)$ (red) and $g^{(2)}_{ii}(\tau)$ (blue) as function of arrival time difference between
two signal photons within the whole phase-matching range operated around 151.2~$^{\circ}$C.}
\label{Narrowg2ssii}
\end{figure}

For different mode numbers $N$, we always have $g^{\left( 2 \right)}\left( 0  \right)=2$, but only for a perfect time resolution. However, realistic detectors suffer from finite time resolution.  For our resonant PDC, the photon has longer temporal wave-packets than the time window of the detectors. Therefore, our measurement provides information about the correlation function with finite temporal resolution, which is different from the usual time-integrated measurements as given for example in Ref.~\cite{ChristNJP2011}. Assuming that the detection window at zero delay point has a window with width $T$, then the measured
$g_{m}^{\left( 2 \right)}\left( 0 \right)$ is given by the convolution of the real $g^{\left( 2 \right)}\left( 0 \right)$ with detector window function.
Concerning the realistic measurement, the time window of our detection system is around 0.5 ns, the $\Delta _{F}$ has magnitude around tens or hundreds of GHz, i.e. the detector window covers several beating periods. Thus, we observe a time-averaging over the beating periods resulting in a measured $g_m^{\left( 2 \right)}\left( 0  \right)$ of
\begin{eqnarray}
g_m^{\left( 2 \right)}\left( 0 \right) & = & 1 + {1 \over N}\underbrace {{1 \over {2T}}\int_{ - T}^T {{{\left| {{g_0}\left( \tau  \right)} \right|}^2}d\tau } }_{ \approx 1} \\ \nonumber
& + & {1 \over {{N^2}}}\underbrace {{1 \over T}\sum\limits_{n = 1}^{N - 1} {\int_{ - T}^T {\left( {N - n} \right)\cos \left[ {n{\Delta _{Fs}}\tau } \right]{{\left| {{g_0}\left( \tau  \right)} \right|}^2}d\tau } } }_{ \approx 0} \\ \nonumber
& \approx & 1 + {1 \over N}.
\label{eq:eq:Norg2m02}
\end{eqnarray}
Consequently, the number of cavity modes can be directly determined from the auto-correlation measurements. Please note that this conclusion is similar for pulsed time-integrated correlations in Ref.~\cite{ChristNJP2011}, but cavity modes are not Schmidt modes. As soon as we use a cw laser to pump, we can not decompose our resonant PDC state to two independent orthonormal bases.

The time-resolved auto-correlation measurement can be used to characterize the spectral longitudinal mode properties of the source.
A measured ${g_m^{\left( 2 \right)}(0)}$ value close to 2 tells that there is only one mode with thermal photon-number distribution. If more cavity modes are involved, APDs cannot distinguish between the different thermal distribution. The measured value is a convolution between each individual thermal photon streams. It causes the small ${g_m^{\left( 2 \right)}(0)}$ value in the finite detector window, which tends towards a Poissonian photon-number distribution after detection. If the measurements go
further away from the auto-correlation time, the time-averaged correlation function drops to its minimum value one as expected.

A measured $g^{(2)}_{ss,ii}(\tau)$ characteristic for unfiltered signal and idler beams is shown in \Fig{Narrowg2ssii}. From the measured value $g_m^{(2)}(0)\approx 1.25$, an effective longitudinal mode number of $N=4$ can be estimated. This is in reasonably good agreement with the simulated output spectrum shown in \Fig{Resonantcluster}(b), where we estimated one predominantly single mode together with several inequivalent modes within the whole phase-matching envelope. Comparing these two auto-correlation curves, it is found that they have different temporal bandwidths, $T _{au_s} \sim$ 12.6~ns and $T _{au_i} \sim$ 9.7~ns, respectively, which are corresponding to the different decay times from signal and idler cavity. Although signal and idler should have the same bandwidth due to the energy conservation and doubly resonant conditions, the auto-correlation still reveals the different cavity behaviors in the resonant waveguide. The reason is that our pump has 200 ns pulse length in order to synchronize the generation and detection systems. This means, the pump still has finite bandwidths around 30 MHz in spectral domain. Besides, the positions of measured $g^{(2)}_{ss,ii}(0)$ slightly shift, just because of imperfect calibrations between two kinds of different detectors (Silicon and InGaAs APDs).

\end{appendix}

\end{document}